\documentclass[final,1p,times]{elsarticle}

\journal{Annals of Physics}
\begin{document}

\begin{frontmatter}
\title{An empirically equivalent random field for the quantized electromagnetic field}
\author{Peter Morgan}
\ead{peter.w.morgan@yale.edu}
\ead[url]{http://pantheon.yale.edu/~pwm22/}
\address{Physics Department, Yale University, New Haven, CT 06520, USA.}
\begin{abstract}
A random field that is empirically equivalent to the quantized electromagnetic field is constructed.
A mapping between the creation and annihilation operator algebras of a random field and of the quantized
electromagnetic field provides a functor between the algebras and the Hilbert spaces generated by the
vacuum states over those algebras.
The functor inevitably does not extend to a functorial relationship between the local algebras generated
by the random field and by the quantized electromagnetic field, but the empirical content provided by
the vacuum state restores an empirical equivalence through the Hilbert spaces.
The isomorphism from one creation and annihilation algebra to the other is not translation invariant because
it depends on mapping positive frequency modes of one helicity to equivalent negative frequency modes, but
the two theories taken independently are presented in equally well-defined and manifestly Lorentz and
translation covariant ways.
\end{abstract}
\end{frontmatter}

\newcommand{\Intd}{{\mathrm{d}}}
\newcommand{\Half}{{\scriptstyle\frac{\scriptstyle 1}{\scriptstyle 2}}}
\newcommand{\me}{{\mathrm{e}}}
\newcommand{\mi}{{\mathrm{i}}}
\newcommand{\abullet}{{\mathsf{a}}}

\section{Introduction}
The construction of a random field that is empirically equivalent to the quantized complex Klein-Gordon field that
was given in \cite{MorganQKGeq} depended on the geometrically unmotivated complex structure of the complex Klein-Gordon
field.
The extension of that approach to the quantized electromagnetic field uses the action of the Hodge dual on the
electromagnetic field as a complex structure in a comparable way.
The mathematics is quite elementary, so, after a brief presentation of the quantized electromagnetic field in Section
\ref{QEM}, it will be presented with relatively little motivation in Section \ref{RandomFieldEQ}, then the more delicate
question of how to interpret the mathematics will be discussed in Section \ref{Discussion}.
The interpretation of the mathematics that is suggested here differs from the interpretation given in~\cite{MorganQKGeq}.

There has been discussion of the similarities and differences between ``random electrodynamics'' and quantum
electrodynamics at least since the 1960s\cite{BoyerA,BoyerB}, however the algebraic formalism used here makes the
comparison, and the establishment of empirical equivalence, much more direct.
The negative frequencies that appear explicitly in the algebraic approach here appear implicitly in random
electrodynamics (which has come to be called ``Stochastic Electrodynamics'' or ``SED''\cite{SED}) as a factor
$\cos{(\vec k\cdot\vec r-\omega t)}$ in the 2-point correlation functions of the random electromagnetic field,
\cite[Eqs. (10), (13), and (14)]{BoyerA}, instead of $\me^{\mi(\vec k\cdot\vec r-\omega t)}$ for the quantized
electromagnetic field, \cite[Eqs. (15), (16), and (17)]{BoyerA}.
A broader analogy with SED is inappropriate, however, because SED's non-field approach to electrons and positrons as
particles rules out any very direct functorial relationship to quantum electrodynamics.

The relationship between random field and quantum field test functions is comparable to the relationship between
the electric and magnetic test functions in \cite[\S V]{BB}.

\section{The quantized electromagnetic field}
\label{QEM}
The quantized electromagnetic field, which is properly speaking an operator-valued distribution, can be constructed
more effectively using test functions (or, as they are called in signal processing, window functions) than is
possible using standard textbook methods.
We use a Schwartz space of test functions $\mathcal{S}$ to smear the operator-valued distributions of the quantized
electromagnetic field, the 2-form $\hat\phi_{\mu\nu}(x)$, to give a complex linear map of test functions into an
algebra of unbounded operators $\mathcal{A}$,
\begin{equation}
  \hat\phi:\mathcal{S}\rightarrow\mathcal{A};f\mapsto \hat\phi_f=\int \hat\phi_{\mu\nu}(x)f^{\mu\nu}(x)\Intd^4x.
\end{equation}
To generate the c-number expectation values associated with operators constructed using $\hat\phi_f$ in the vacuum state,
we write $\hat\phi_f$ as the sum of abstract creation and annihilation operators, $\hat\phi_f=a^{\,}_{f^*}+a_f^\dagger$,
where $a^{\,}_{f^*}$ and $a_f^\dagger$ are engineered both to be complex linear by introducing the local real-space
complex conjugation $f^*(x)=\left[f(x)\right]^*$, and write the positive semi-definite commutator
$[a^{\,}_f,a_g^\dagger]$ as
\begin{equation}\label{EMCommutator}
  [a^{\,}_f,a_g^\dagger]=(f,g)=-\hbar\int k^\alpha\tilde f_{\alpha\mu}^*(k) k^\beta\tilde g_\beta{}^\mu(k)
                                    2\pi\delta(k_\nu k^\nu)\theta(k_0)\frac{\Intd^4k}{(2\pi)^4},
\end{equation}
then we introduce a vacuum vector $\left|0\right>$, on which annihilation operators have a trivial action
$a^{\,}_f\left|0\right>=0$, which allows us to define a complex linear vacuum state
$\omega_0:\mathcal{A}\rightarrow\mathbb{C};\omega_0(\hat A)=\left<0\right|\hat A\left|0\right>$,
and hence to use the GNS-construction of a Hilbert space \cite[\S III.2.2]{Haag}.
The commutator $[\hat\phi_f,\hat\phi_g]=[a^{\ }_{f^*},a_g^\dagger]-[a^{\ }_{g^*},a_f^\dagger]$ is trivial
if the test functions $f$ and $g$ have space-like separated supports, but it is nontrivial in general.

The algebraic structure and the trivial action of the annihilation operators on the vacuum state generates the
c-number expectation value associated with operators that are constructed as sums and products of creation and
annihilation operators.
The same algebraic structure, $[a^{\,}_f,a_g^\dagger]=(f,g)$, works for any free quantum field, with different space-time
properties encoded in different forms of the (positive semi-definite) inner product $(f,g)$.
The form given above for the quantized electromagnetic field is derived, for example, in \cite[Eq. (3.27)]{MenikoffSharp}.

\subsection{The electromagnetic potential}
In terms of test functions, an account that uses the electromagnetic potential is largely equivalent to the above.
For the electromagnetic potential operator-valued distribution smeared by a test function $u^\rho(x)$,
$\hat A_u=\int_M \hat A_\rho(x)u^{\rho*}(x)\Intd^4x$, to be an observable that is invariant under
$U(1)$ gauge transformations $\hat A_\rho(x)\rightarrow\hat A_\rho(x)-\partial_\rho\alpha(x)$, we require that
$\int_M \partial_\rho\alpha(x)u^{\rho*}(x)\Intd^4x$ must be zero for all scalar functions $\alpha(x)$.
Integrating by parts over a region $\Omega$ in Minkowski space, we obtain, in terms of differential forms,
\[
  \int_\Omega d\alpha\wedge(\star u^*)=\int_{\partial\Omega}\alpha\wedge(\star u^*)-\int_\Omega \alpha\wedge(d\!\star\! u^*),
\]
which will be zero for large enough $\Omega$, and hence for the whole of Minkowski space, for any smooth test function
$u^\rho(x)$ that has compact support and is divergence-free, $d\!\star\! u=0$.
If we constrain the gauge transformation function $\alpha(x)$ not to increase faster than polynomially with increasing
distance in any direction, it will be enough for the test function $u^\rho(x)$ to be Schwartz and divergence-free.

\newcommand\EMPa{{\mathbf{\scriptstyle a}}}
The divergence-free condition on $u^\rho(x)$ ensures that the commutator for creation and annihilation operators
associated with the electromagnetic potential $\hat A_u=\EMPa^{\,}_{u^*}+\EMPa^{\dagger}_u$,
\[
  [\EMPa^{\,}_u,\EMPa^\dagger_v]=-\hbar\int \tilde u^*_\rho(k)\tilde v^\rho(k)
                                                  2\pi\delta(k_\nu k^\nu)\theta(k_0)\frac{\Intd^4k}{(2\pi)^4},
\]
is positive semi-definite (which is required for us to be able to construct a vacuum sector Hilbert space), and
that because $\delta u=\delta v=0$ we can construct, on Minkowski space, $u=\delta U$, $v=\delta V$, where
$U$ and $V$ are bivector potentials for the electromagnetic potential test functions $u$ and $v$.
In terms of $U$ and $V$, we can write $a^{\,}_U=\EMPa^{\,}_{\delta U}$, $a_V^\dagger=\EMPa^\dagger_{\delta V}$, which
satisfy the commutator (\ref{EMCommutator}).
Consequently, turning around the usual relationship because we are working with test functions instead of directly with
quantum fields, we can regard test functions for the electromagnetic field as potentials for test functions for the
electromagnetic potential.

The restriction that electromagnetic potential test functions must have compact support (or that gauge
transformations must be constrained if the electromagnetic potential test functions are taken to be Schwartz)
means that electromagnetic potential observables are less general than electromagnetic field observables if
electromagnetic field test functions are taken to be Schwartz (as is most commonly assumed), or equivalent
if electromagnetic field test functions are taken to be smooth and to have compact support.

\section{A random field equivalent}
\label{RandomFieldEQ}
The construction of a Klein-Gordon random field that is empirically equivalent to the complex Klein-Gordon
field~\cite{MorganQKGeq} depended on the Lorentz invariant projection to positive and negative frequency components
and the Poincar\'e invariant projection to real and imaginary components.
For the quantized electromagnetic field we use the Poincar\'e invariant projection to positive and negative helicity
components, instead of the projection to real and imaginary components, to construct an involution
\begin{equation}
  {}^\bullet:\mathcal{S}\rightarrow \mathcal{S};f\mapsto f^\bullet,\qquad
     \widetilde{f^\bullet}(k)=\Half(1+i\star)\tilde f(k)+\Half(1-i\star)\tilde f(-k),\qquad f^{\bullet\bullet}=f,
\end{equation}
where for the Hodge dual $\star$, $(\star\hspace{-0.25em}f)_{\mu\nu}=\Half\epsilon_{\mu\nu}{}^{\alpha\beta}f_{\alpha\beta}$,
$\star\hspace{-0.4em}\star\hspace{-0.4em}f=-f$, and we have omitted indices from the above equation.
In terms of $f^\bullet$, we can define modified annihilation operators, $\abullet_f=a^{\ }_{f^\bullet}$, for which, with
the corresponding creation operator, we can compute the commutator
\begin{eqnarray}
  [\abullet^{\ }_f,\abullet_g^\dagger]&=&[a^{\ }_{f^\bullet},a_{g^\bullet}^\dagger]=(f^\bullet,g^\bullet)\cr
             &&\hspace{-4em}=-\hbar\int\frac{\Intd^4k}{(2\pi)^4}2\pi\delta(k_\nu k^\nu)\theta(k_0)k^\alpha k^\beta
                            \widetilde{f^\bullet}^*_{\alpha\mu}(k)\widetilde{g^\bullet}_\beta^{\hspace{0.6em}\mu}(k)\cr
             &&\hspace{-4em}=-\hbar\int\frac{\Intd^4k}{(2\pi)^4}2\pi\delta(k_\nu k^\nu)\theta(k_0)k^\alpha k^\beta\cr
             &&\hspace{2em}\times\Bigl[\Half(1+i\star)\tilde f(k)+\Half(1-i\star)\tilde f(-k)\Bigr]^*_{\alpha\mu}
                                 \Bigl[\Half(1+i\star)\tilde g(k)+\Half(1-i\star)\tilde g(-k)\Bigr]_\beta^{\hspace{0.6em}\mu}\cr
             &&\hspace{-4em}=-\hbar\int\frac{\Intd^4k}{(2\pi)^4}2\pi\delta(k_\nu k^\nu)\theta(k_0)k^\alpha k^\beta\Bigg(
                            \Bigl[\Half(1+i\star)\tilde f(k)\Bigr]^*_{\alpha\mu}
                            \Bigl[\Half(1+i\star)\tilde g(k)\Bigr]_\beta^{\hspace{0.6em}\mu}\cr
             &&\hspace{10em}+\Bigl[\Half(1-i\star)\tilde f(-k)\Bigr]^*_{\alpha\mu}
                             \Bigl[\Half(1-i\star)\tilde g(-k)\Bigr]_\beta^{\hspace{0.6em}\mu}\Bigg),
\end{eqnarray}
which is, as is required for a fundamental field, invariant under translations of the test functions $f$ and $g$, even
though the transformation $f\mapsto f^\bullet$ is not.
To construct an observable random field $\hat\chi_f$ that satisfies the trivial commutator
$[\hat\chi_f,\hat\chi_g]=0$, we define 
\[
  \hat\chi_f=\abullet^{\ }_{f^*}+\abullet_f^\dagger=a^{\ }_{f^{*\bullet}}+a_{f^\bullet}^\dagger
            \qquad\not=\ a^{\ }_{f^{\bullet*}}+a_{f^\bullet}^\dagger=\hat\phi_{f^\bullet},
\]
for which $[\hat\chi_f,\hat\chi_g]=[\abullet^{\ }_{f^*},\abullet_g^\dagger]-[\abullet^{\ }_{g^*},\abullet_f^\dagger]$;
we find that 
\begin{eqnarray}
  [\abullet^{\ }_{f^*},\abullet_g^\dagger]&=&[a^{\ }_{f^{*\bullet}},a_{g^\bullet}^\dagger]=(f^{*\bullet},g^\bullet)\cr
             &&\hspace{-4em}=-\hbar\int\frac{\Intd^4k}{(2\pi)^4}2\pi\delta(k_\nu k^\nu)\theta(k_0)k^\alpha k^\beta\Bigg(
                            \Bigl[\Half(1+i\star)\widetilde{f^*}(k)\Bigr]^*_{\alpha\mu}
                            \Bigl[\Half(1+i\star)\tilde g(k)\Bigr]_\beta^{\hspace{0.6em}\mu}\cr
             &&\hspace{10em}+\Bigl[\Half(1-i\star)\widetilde{f^*}(-k)\Bigr]^*_{\alpha\mu}
                            \Bigl[\Half(1-i\star)\tilde g(-k)\Bigr]_\beta^{\hspace{0.6em}\mu}\Bigg)\cr
             &&\hspace{-4em}=-\hbar\int\frac{\Intd^4k}{(2\pi)^4}2\pi\delta(k_\nu k^\nu)\theta(k_0)k^\alpha k^\beta\Bigg(
                            \Bigl[\Half(1-i\star)\tilde f(-k)\Bigr]_{\alpha\mu}
                            \Bigl[\Half(1+i\star)\tilde g(k)\Bigr]_\beta^{\hspace{0.6em}\mu}\cr
             &&\hspace{10em}+\Bigl[\Half(1+i\star)\tilde f(k)\Bigr]_{\alpha\mu}
                            \Bigl[\Half(1-i\star)\tilde g(-k)\Bigr]_\beta^{\hspace{0.6em}\mu}\Bigg),
\end{eqnarray}
which is symmetric in $f$ and $g$, so that
$[\hat\chi_f,\hat\chi_g]=[\abullet^{\ }_{f^*},\abullet_g^\dagger]-[\abullet^{\ }_{g^*},\abullet_f^\dagger]=0$ for any
test functions $f$ and $g$, not only when they have space-like separated supports.
The universal commutativity of $\hat\chi_f$ for all test functions depends on the fact that $f^{*\bullet}\not=f^{\bullet*}$.

There is a similar freedom of description in quantum field theory that is only occasionally noted, which can be
presented using the relatively trivial map $f\mapsto f^\circ=\me^{\mi\alpha}f$, $\mathring{a}^{\,}_f=a^{\,}_{f^\circ}$.
This gives us an alternative quantum field,
\begin{equation}
  \hat\phi'_f=\mathring{a}_{f^*}+\mathring{a}_f^\dagger=a^{\,}_{\me^{\mi\alpha}(f^*)}+a_{\me^{\mi\alpha}f}^\dagger
                                 =\me^{-\mi\alpha}a^{\,}_{f^*}+\me^{\mi\alpha}a_{f}^\dagger
                   \qquad    \not=\hat\phi_{\me^{\mi\alpha}f}=\me^{\mi\alpha}\hat\phi_f,
\end{equation}
which satisfies microcausality in the same way as $\hat\phi_f$, $[\hat\phi'_f,\hat\phi'_g]=[\hat\phi_f,\hat\phi_g]$.
Weinberg notes, for example, that we cannot have both $\hat\phi_f$ and $\hat\phi'_f$ in the same
theory~\cite[p203]{Weinberg}, so we have to make a conventional choice of phase.
Equally, to transform to a random field formalism we have to make a conventional choice of phases and use
only that choice.

\section{Discussion}\label{Discussion}
We have shown that there exist equally effective alternative presentations of the Hilbert space of states, either
in terms of random electromagnetic field observables or in terms of quantized electromagnetic field observables.
Exactly the same Hilbert space is generated by the action of the creation operators $\abullet_f^\dagger$ on the vacuum vector
as is generated by the action of the creation operators $a_f^\dagger$ on the vacuum vector, even
though the algebra of local observables that is generated by $a^{\ }_{f^*}+a^\dagger_f$ is significantly different from
the algebra of local observables that is generated by $\abullet^{\ }_{f^*}+\abullet^\dagger_f$.
The algebra of observables that is used in quantum optics and in other Physics is routinely extended, however, by
the nonlocal vacuum projection operator $\hat V=\left|0\right>\left<0\right|$, because this or an equivalent extension
of the algebra of creation and annihilation operators is necessary to allow the elementary construction of observables
that have discrete eigenvalues that can be used to model the statistics of discrete measurement events.
For example, a transition event observable between a normalized vector state
$\left|\Psi\right>=\hat\Psi^\dagger\left|0\right>$
(where $\hat\Psi^\dagger$ is, in the elementary case, a multinomial in some set of field operators)
and a normalized vector state $\left|\Psi'\right>$ can be presented as the squared modulus of an amplitude
\[
  \left|\left<\Psi\right|\!\left.\Psi'\right>\right|^2=\left<\Psi\right|\hat V_{\hat\Psi'}\left|\Psi\right>,
\]
which can be understood as a measurement of the observable $\hat V_{\hat\Psi'}=\hat{\Psi'}^\dagger\hat V\hat\Psi'$ in
the state $\omega_{\hat\Psi}(\hat A)=\left<\Psi\right|\hat A\left|\Psi\right>$. 
$\hat V_{\hat\Psi'}$ is nonlocal because of the presence of the vacuum projection operator, insofar as the commutator
$[\hat V_{\hat\phi_{f}},\hat V_{\hat\phi_{g}}]$ is generally nonzero when the functions $f(x)$ and $g(x)$ have
space-like separated supports.
With the introduction of the vacuum projection operator, the space of noncommuting, nonlocal observables generated
by $\hat\phi_f$ and at least one $\hat V$ is the same as the space of noncommuting, nonlocal observables generated
by $\hat\chi_f$ and at least one $\hat V$.
It is notable that it is only with the introduction of the vacuum projection operator that the algebra generated
by $\{\hat\chi_f,\hat V\}$ becomes noncommutative, but then everything is the same as the algebra generated by
$\{\hat\phi_f,\hat V\}$.

At a higher mathematical level than that of the rest of this article, the mapping $\abullet^{\ }_{f}=a^{\ }_{f^\bullet}$
provides us with a category-theoretic functor between the algebras of creation and annihilation operators and the Hilbert
spaces that the vacuum state allows us to construct, that does not extend, however, to a functor between the algebras of
local observables that are generated by $\hat\chi_f$ and $\hat\phi_f$.
The functorial relationship also does not extend straightforwardly to the less invariant algebraic structures that are
associated with phase spaces on space-like hypersurfaces.

It is usual in Physics to keep an apparatus as isolated as possible from all known interference, but any
interference that cannot be removed is, as far as possible, measured and corrected for.
In experimental reports, details are given of all the interference that was eliminated or that was measured
and corrected for, however measurement and correction is generally taken not to be possible for the effects of
quantum fluctuations.
This is justified because of the apparent universality of quantum fluctuations, in particular extending to all
measurement apparatus, however the elementary mathematics given in~\cite{MorganQKGeq} for the quantized
complex Klein-Gordon field, and here for the physically more relevant quantized electromagnetic field,
allows quantum fluctuations to be corrected for by reporting what experimental results \emph{would}
be observed by \emph{ideal} local measurements represented by a random field.
At the same time as we \emph{can} report what measurement results would be if they were not deeply influenced by
quantum fluctuations, however, the reality of experiments is that there are incompatibilities between real measurements,
which we typically model using nonlocal projection operators, and which typically are statistics of engineered
thermodynamic transition events, making quantized electromagnetism often more natural.

When the effects of quantum fluctuations on measurements can be ignored, we can work with the expected values of
the observable $\hat\phi_{\mu\nu}(x)$ as if it is the classical electromagnetic field, as an effective mean field theory.
When significant effects are caused by quantum fluctuations, however, so that a mean field theory is not accurate enough,
we have to accommodate the consequences for measurement results in some way.
In the quantum field theory that is generated by $\hat\phi_{\mu\nu}(x)$, the effects of quantum fluctuations are modeled
by the commutator (\ref{EMCommutator}), with the real part of the commutator determining a
minimum level of fluctuations associated with the elementary observable $\hat\phi_f$ in the vacuum state and
with the imaginary part of the commutator determining a minimum level of incompatibility between
time-like separated measurements.
In the equivalent random field theory, using $\hat\chi_{\mu\nu}(x)$, by taking the positive and negative helicity
components to be of opposite frequency the incompatibility between time-like separated measurements is eliminated,
while preserving the same level of fluctuations of the equivalent components.

\end{document}